\PassOptionsToPackage{pdfpagelabels=true}{hyperref} 

\documentclass[fleqn,usenatbib]{mnras}

\usepackage{newtxtext,newtxmath}
\usepackage[utf8]{inputenc}

\usepackage{natbib}
\usepackage{hypernat} 

\usepackage[T1]{fontenc}
\usepackage{ae,aecompl}

\usepackage{graphicx}	
\usepackage{amsmath}	
\usepackage{newtxtext,newtxmath}
\usepackage{multirow}

\title[Effect of fragmentation on super-Earth formation]{The ``Breaking The Chains'' migration model for super-Earths formation: the effect of collisional fragmentation}

\author[L. Esteves et al.]{
Leandro Esteves,$^{1}$\thanks{E-mail: leandro.esteves@unesp.br}
André Izidoro,$^{2,1}$\thanks{E-mail: izidoro.costa@gmail.com}
Bertram Bitsch,$^{3}$
Seth A. Jacobson,$^{5}$
\newauthor
Sean N. Raymond,$^{4}$ Rogerio Deienno,$^{6}$
and Othon C. Winter$^{1}$
\\
$^{1}$UNESP, Universidade Estadual Paulista, Grupo de Dinâmica Orbital e Planetologia, Guaratinguetá, CEP 12516-410, SP, Brazil\\
$^{2}$Department of Earth, Environmental and Planetary Sciences, MS 126, Rice
University, Houston, TX 77005, USA\\
$^{3}$Max-Planck-Institut für Astronomie, Königstuhl 17, 69117 Heidelberg, Germany\\
$^{4}$Laboratoire d’astrophysique de Bordeaux, Univ. Bordeaux, CNRS, B18N, allée Geoffroy Saint-Hilaire, 33615 Pessac, France\\
$^{5}$Department of Earth and Environmental Sciences, Michigan State University, East Lansing, MI 48824, USA \\
$^{6}$Southwest Research Institute, 1050 Walnut St. Suite 300, Boulder, CO 80302, USA
}

\date{Accepted XXX. Received YYY; in original form ZZZ}

\pubyear{2021}

\begin{document}
\label{firstpage}
\pagerange{\pageref{firstpage}--\pageref{lastpage}}
\maketitle

\begin{abstract}
Planets between 1-4 Earth radii with orbital periods <100 days are strikingly common. The migration model proposes that super-Earths migrate inwards and pile up at the disk inner edge in chains of mean motion resonances. After gas disk dispersal, simulations show that super-Earth’s gravitational interactions can naturally break their resonant configuration leading to a late phase of giant impacts. The instability phase is key to matching the orbital spacing of observed systems. Yet, most previous simulations have modelled collisions as perfect accretion events, ignoring fragmentation. In this work, we investigate the impact of imperfect accretion on the breaking the chains scenario. We performed N-body simulations starting from distributions of planetary embryos and modelling the effects of pebble accretion and migration in the gas disk. Our simulations also follow the long-term dynamical evolution of super-Earths after the gas disk dissipation. We compared the results of simulations where collisions are treated as perfect merging events with those where imperfect accretion and fragmentation are allowed. We concluded that the perfect accretion is a suitable approximation in this regime, from a dynamical point of view. Although fragmentation events are common,  only $\sim$10\% of the system mass is fragmented during a typical “late instability phase”, with fragments being mostly reacreted by surviving planets.  This limited total mass in fragments proved to be insufficient to alter qualitatively the final system dynamical configuration -- e.g. promote strong dynamical friction or residual migration -- compared to simulations where fragmentation is neglected.
\end{abstract}




\section{Introduction}

Planets with sizes between those of Earth (1~R$_{\oplus}$) and Neptune (4~R$_{\oplus}$) -- usually referred to as super-Earths (or mini-Neptunes) -- are remarkably common around other stars. At least $\sim$30-50\% of the FGK-spectral type stars host super-Earths with orbital periods shorter than 100 days \citep{mayoretal11,howard12,fressin2013,marcy2014,muldersetal18}. Mercury -- the innermost solar system planet -- has an orbital period of 88 days. Most observed super-Earths orbit their respective stars at orbits relatively closer than Mercury is to the Sun. The high level of stellar irradiation that these planets receive from their parent stars has led these planets to be called ``hot'' super-Earths (or hot mini-Neptunes). Given their apparent ordinariness and astonishing orbital configuration compared to solar system planets, the origins of hot super-Earths is today an intense area of research. Models for the formation of hot super-Earths come in two flavours. In one school of thought hot super-Earths are believed to have reached the innermost regions of the disk via gas-driven migration as they grow during the gas disk phase~\cite[e.g.][]{terquempapaloizou07,raymond08,idalin098,colemannelson14,cossouetal14,izidoro2017breaking}. In the second school of thought,  super-Earths are believed to have formed  ``in-situ'' and without experiencing gas-driven radial  migration~\citep[e.g][]{raymond08,hansen14,leeetal14,ogiharaetal15}.

In this work we model the formation of hot super-Earths systems in the context of the migration model. For a detailed discussion on why super-Earths are very unlikely to have completely avoided migration and formed  ``in-situ'' see \cite{izidoro2019formation}.

The migration model for the origins of hot super-Earths proposes that super-Earth migrated inwards during the gas disk phase due to the planet-disk gravitational interactions \citep[][]{terquempapaloizou07,raymond08,mcneilnelson10,idalin10}. Super-Earths  migrating inwards are not necessarily engulfed by the young star but most likely trapped at the gas disk inner edge ~\citep{massetetal06,romanovalovelace06,flocketal19}. This dynamical evolution tend to produce long-chains of planets locked in  mean motion resonances  anchored at the disk inner edge ~\citep{ogiharaida09,cossouetal14,izidoro2017breaking,ogiharaetal18,lambrechtsetal19,izidoro2019formation,carreraetal19}. Planet-disk gravitational interaction effects damp planets' orbital inclinations and eccentricities \citep{cresswell2008three,bitschkley10,bitschkley11,fendykenelson14}. As the gas disk dissipates, damping of orbital inclination and eccentricities becomes less efficient and eventually vanish when the gas is fully dispersed. Numerical simulations show that the subsequent evolution of resonant chains bifurcates in two categories. In the first one, super-Earth systems remain dynamically stable preserving their pristine resonance state for billion of years~\citep{estevesetal20}. In the second category, resonant chains become  dynamically unstable -- typically at the very end or just after the gas disk dispersal ($\lesssim$100Myr). Late dynamical instabilities break resonances and lead to orbital crossing and giant impacts~\citep{izidoro2017breaking,izidoro2019formation}. Systems that have experienced the late instability phase are typically refereed as ``unstable''. These planetary systems tend to be dynamically spread (non-resonant) with planets having relatively larger mutual orbital inclinations and eccentricities. On the contrary,  ``stable'' systems tend to have planets on very low eccentricity and inclination. 

Numerical simulations show that the migration model matches fairly well the period ratio distribution of observed super-Earths  when  $\gtrsim$95-99\% of resonant chains become dynamically unstable  \citep{izidoro2017breaking,izidoro2019formation} and $\lesssim$5-1\% remain stable. We refer to this as the ``Breaking the chain'' scenario (BTC). This model also predicts that most observed planetary systems showing single transit super-Earths are in fact multi-planet systems (N$\geq$2). About $\sim$80\% of the observed stars hosting hot super-Earths show a single transiting planet, whereas only $\sim$20\% of the systems comes with multiple transiting planets. This result is known as the  Kepler's Dichotomy \citep{johansen2012can}. The BTC scenario suggests that the Kepler's dichotomy is a consequence of observational bias ~\citep{izidoro2017breaking,izidoro2019formation} and that most single transiting planets are not truly single. Simulated transit observations mixing a fraction of  $<$5\% stable and $>$95\% unstable systems (fractions that produce a good match to the observed period ratio distribution of super-Earths) produce also a decent match to the Kepler planet multiplicity distribution and dichotomy~\citep{izidoro2017breaking,izidoro2019formation}. The low-mutual orbital inclination of planets in stable systems tend to favour the detection of multiple transiting planets if the observer is well-aligned with the planets' (``common'') orbital plane.  Unstable systems, in contrast, tend to have  planets with relatively larger mutual orbital inclinations which make them more likely to be missed by transit observations. This naturally enhances  the number systems showing single transiting planets.

Although the BTC scenario provides an appealing framework to explain the origin of hot super-Earth system it remains elusive how the simplistic treatment of impacts adopted in previous simulations influences the main results of the model. So far, all simulations in the context of the BTC scenario were performed modelling  collisions as perfect merging events where mass and linear momentum are conserved~\citep{izidoro2017breaking,izidoro2019formation}. Collisional fragmentation have been proposed to lead to self-destruction of hot super-Earths systems~\citep{volkgladman15}. This hypothesis seems to conflict with the results of numerical simulations showing successful planetary growth in regions as close as 0.04~au from the star~\citep{wallace17}. One real possibility, however, is that fragments produced via impacts have the chance to interact with the final planets and extract them from mean motion resonances and/or damp their orbital eccentricity and inclinations~\citep[e.g.][]{chatterjee2015planetesimal}. These effects could strongly alter the final dynamical architecture  of planetary systems (e.g. masses, resonant configuration, number of planets, etc.) and how the migration model matches observations. The main goal of this paper is to revisit the BTC scenario using a more realistic treatment for impacts.  We use N-body numerical simulations to model planetary growth via pebble accretion, gas-driven migration, and the long-term evolution of hot super-Earths. This work builds on previous studies~\citep{izidoro2019formation,lambrechtsetal19,bitschetal19}. We compare the results of simulations where we model collisions using semi-analytical prescriptions  calibrated from the output  of hydrodynamical simulations~\citep{leinhardt2012,stewart12} with the results of simulations where collisions are modelled as as perfect merging events. We discuss the importance of considering imperfect accretion when modelling the formation of hot super-Earths. 
 
This paper is structured as follows: In \S \ref{sec:methods} we present our methods, the collision treatment algorithm and simulations set-up. In \S \ref{sec:simulations} we show the results of representative simulations. In  \S \ref{sec:stat_analysis} we present a statistical analysis comparing the results of simulations where the effects of collisional fragmentation are accounted for and those where it is neglected. In \S \ref{sec:discussion} we discuss about the dust production and mantle-erosive collisions in our simulations. Finally, in \S \ref{sec:conclusion} we discuss our model, results and main conclusions.

\section{Methods}\label{sec:methods}

Our N-body simulations were performed using a new version of the \textsc{Mercury-Flintstone} code \citep{izidoro2019formation,bitschetal19}. \textsc{Mercury-Flintstone} is a modified version of \textsc{Mercury} N-body integrator package \citep{chambers1999hybrid} that includes routines dedicated to model gas-disk evolution, planet migration, gas tidal damping of inclination/eccentricity, gas-assisted pebble accretion and gas accretion \citep[for a detailed description of these routines see][]{bitschetal19,izidoro2019formation}. In this work, we have added a new feature to \textsc{Mercury-Flintstone}. We have upgraded the simplistic treatment of collisions used in the code -- where collisions are always assumed to result in perfect merging events. We have implemented into our code a prescription to resolve collisions based on the outcome of impact hydrodynamical simulations~\citep{leinhardt2012}. Collisions in \textsc{Mercury-Flintstone} can now result in bouncing, fragmentation and accretion. All our simulations were performed using the hybrid-sympletic integrator~\citep{chambers1999hybrid}.

\subsection{Collision Model}\label{collision}

\textsc{Mercury-Flintstone} uses the same algorithm of the regular \textsc{Mercury} to identify collisions. At every timestep it checks if the distance between any two bodies is smaller than the sum of their physical radius. When this condition is matched, a collision is flagged to be resolved.  One of the caveats of this approach is that a collision may be identified  when colliding bodies have already largely overlapping physical radii. In this situation, to properly resolve the collision, we need to restore the dynamical state of the colliding bodies to the very moment  when they ``touch'' each other.  We account for this issue by integrating the motion of the colliding bodies back in time. We use a three-body approximation (sun and colliding bodies) and the Bulirsch-Stoer integration algorithm with accuracy parameter set to $10^{-15}$, which ensures a small timestep.  We stop this reversal integration at the first timestep when the distance between the bodies becomes larger than their combined physical radius.  We use the dynamical state of colliding bodies at this specific instant to resolve the collisions as described below.

Following \cite{leinhardt2012}, we define the more massive body involved in a collision as ``target'' (thereafter use the subscript t) while the less massive planetary body is referred as  ``projectile'' (thereafter we use the subscript p). $M_{\text{t}}$ and $M_{\text{p}}$ represent the masses of the target and projectile, respectively. $R_{\text{t}}$ and $R_{\text{p}}$ are their respective physical radius. $V_{\text{imp}}$ is their impact velocity, $Q$ is the impact energy, $b$ is the impact parameter given by $b = \textrm{sin}(\theta )$, where $\theta$ is the impact angle. $b_{\text{crit}}$ represents the critical impact parameter that sets the threshold between grazing ($b > b_{\text{crit}}$) and non-grazing ($b < b_{\text{crit}}$) collisions. $M_{\text{int}}$ is the interacting total mass, $Q'^*_{\text{RD}}$ is the critical impact energy, and $Q^{\dagger*}_{\text{RD}}$ is the critical impact energy of the reverse-impact (we refer the reader to \cite{leinhardt2012} for detailed description of these quantities).

Following \cite{leinhardt2012}, we express the outcome of a collision in terms of: i) the largest (most massive) remnant object; ii) the second-largest remnant object; and iii) a population of relatively smaller fragments. We also classify impacts in different regimes: i) perfect merge; ii) super-catastrophic; iii) partial erosion; iv) partial accretion; vi) erosive hit-and-run; vii) Graze-and-merge; and viii) pure hit-and-run. For each impact category we describe how we calculate the masses of the largest and second largest remnants, and respective  smaller fragments. We refer to the largest and second largest remnant using the subscripts $\text{LR}$ and SLR, respectively (e.g. $M_{\text{lr}}$ is largest remnant mass). In our model, mass is always conserved when we resolve collisions, i.e., $M_{\text{t}}+M_{\text{p}}=M_{\text{lr}}+M_{\text{slr}}+M_{\text{frag}}=M_{\text{tot}}$, where $M_{\text{frag}}$ represents the total mass in fragments. We start by describing impact regimes within the non-grazing category ($b < b_{\text{crit}}$).

\begin{enumerate}
  {\bf   \item Perfect merge regime:} if $V_{\text{imp}} < V_{\alpha,\text{esc}}$, where $V_{\alpha,\text{esc}}$ is the surface escape velocity of the interacting mass ($M_{\text{int}}$), given by: \label{r1}
\begin{equation}
V_{\alpha,\text{esc}} = \sqrt{2GM_{\text{int}}/(R_{\text{t}}+R_{\text{p}})}.
\end{equation}
In the perfect merge regime, $M_{\text{lr}}=M_{\text{t}}+M_{\text{p}}$, $M_{\text{slr}}=0$, and  $M_{\text{frag}}=0$. 

{\bf   \item Super-catastrophic regime:} if $V_{\text{imp}} \geq V_{\text{sc}}$, where $V_{\text{sc}}$ is the super-catastrophic critical velocity expressed in terms of the super-catastrophic disruption energy $Q_{\text{sc}} = 1.8Q'^*_{\text{RD}}$, collision combined mass $M_{\text{tot}}=M_{\text{t}}+M_{\text{p}}$ and reduced mass $\mu = M_{\text{t}}M_{\text{p}}/M_{\text{tot}}$, defined as \label{r2}
\begin{equation}
V_{\text{sc}} = \sqrt{2Q_{\text{sc}}M_{\text{tot}}/\mu}.
\end{equation}

The super-catastrophic regime corresponds to the case where fragmentation takes place and the largest remnant receives less than 10\% of the impact combined mass. The mass of the largest remnant is calculated as
\begin{equation}
M_{\text{lr}} = 0.1M_{\text{tot}}\left ( \frac{Q}{1.8Q'^*_{\text{RD}}} \right )^{-3/2}. 
\end{equation}

{\bf     \item Partial erosion regime:} if $V_{\text{imp}} \geq V_{\text{ero}}$, where $V_{\text{ero}}$ is the erosive critical velocity expressed as 
\begin{equation}
V_{\text{ero}} = \sqrt{2Q_{\text{ero}}M_{\text{tot}}/\mu},
\end{equation}
with the critical erosive energy set $Q_{\text{ero}} = Q'^*_{\text{RD}}\left (\frac{2M_p}{M_{\text{tot}}}\right )$. The mass of the largest remnant is given as
    \begin{equation}
    M_{\text{lr}} = M_{\text{tot}}\left (1- \frac{Q}{2Q'^*_{\text{RD}}} \right ).;
    \end{equation}

  {\bf   \item Partial accretion regime:} if $b < b_{\text{crit}}$ and $V_{\text{imp}} < V_{\text{ero}}$. \\
    The transition between the erosion and accretion regimes occurs when the largest remnant mass becomes smaller than the initial target mass. As in the previous regime,
        \begin{equation}
    M_{\text{lr}} = M_{\text{tot}}\left (1- \frac{Q}{2Q'^*_{\text{RD}}} \right );
    \end{equation}

For grazing impacts ($b > b_{\text{crit}}$) the following regimes can be defined:

   {\bf  \item Super-catastrophic hit-and-run regime:} if $V_{\text{imp}} \geq V^{\dagger}_{\text{sc}}$, where $V^{\dagger}_{\text{sc}}$ is the super-catastrophic critical  velocity of the reverse impact defined as
\begin{equation}
V^{\dagger}_{\text{sc}} = \sqrt{2Q^{\dagger}_{\text{sc}}M_{\text{int}}/\mu_{\text{int}}}.
\end{equation}
$\mu_{\text{int}}$ is the interacting reduced mass. $Q^{\dagger}_{\text{sc}} = 1.8Q^{\dagger*}_{\text{RD}}$ defines the critical impact energy for an impact where the largest remnant from the projectile receives less than 10\% of the interacting total mass ($M_{\text{int}}$). In this case, $M_{\text{lr}} = M_t$ (fragmentation occurs only for the projectile);

  {\bf   \item Erosive hit-and-run regime:} if $V_{\text{imp}} \geq V^{\dagger}_{\text{ero}}$, where $V^{\dagger}_{\text{ero}}$ is the erosive critical velocity of the reverse impact given by
\begin{equation}
V^{\dagger}_{\text{ero}} = \sqrt{2Q^{\dagger}_{\text{ero}}M_{\text{int}}/\mu},
\end{equation}
    where the critical erosive impact energy is $Q^{\dagger}_{\text{ero}} = Q^{\dagger*}_{\text{RD}}\left (\frac{2M_{\text{p,int}}}{M_{\text{int}}}\right )$. $M_{\text{p,int}}$ is the projectile mass that effectively interacts with the target. In this case $M_{\text{lr}}=M_{\text{t}}$ and 
    \begin{equation}
    M_{\text{slr}} = M_{\text{int}}\left (1- \frac{Q^{\dagger}}{2Q^{\dagger*}_{RD}} \right ),
    \end{equation}
    The fragments total mass is $M_{\text{frag}} = M_{\text{tot}}-M_{\text{lr}}-M_{\text{slr}}$. 

 {\bf    \item Graze-and-merge regime:} if $V_{\text{imp}} < V_{\text{hr}}$, where $V_{\text{hr}}$ is the threshold velocity defining the boundary between the hit-and-run and graze-and-merge regimes. Following \cite{kokubogenda10} it is defined as:
\begin{equation}
\frac{V_{\text{hr}}}{V_{\text{esc}}}= c_1 \zeta^2 (1-b)^{5/2}+c_2\zeta^2 + c_3(1-b)^{5/2}+c_4,
\end{equation}
    where $\zeta = (M_{\text{t}} - M_{\text{p}})/M_{\text{tot}}$. All coefficients $c_{\text{i}}$ are given in~\cite{gendakokuboida12}. $V_{\text{esc}}$ is the surface escape velocity of the combined mass of the colliding bodies. As in the perfect merge regime, $M_{\text{lr}}=M_{\text{t}}+M_{\text{p}}$, $M_{\text{slr}}=0$.

{\bf     \item Pure hit-and-run regime:} if $V_{\text{imp}} > V_{\text{hr}}$. In this case, the colliding bodies conserve their original masses.\\
\end{enumerate}

 We can now define the mass of the second largest remnant in regimes ii), iii), iv), and v). For these regimes, $M_{\text{slr}}$ is calculated as \cite{leinhardt2012} via the following equation: 
\begin{equation}
    M_{\text{slr}} = M_{\text{tot}}\frac{(3- \beta )\left (1-N_{\text{lr}}\frac{M_{\text{lr}}}{M_{\text{tot}}}  \right )}{N_{\text{slr}}\beta },
\label{eq:slr}
\end{equation}
\noindent where $\beta$ is the power-law slope of the mass/size distribution of fragments obtained via fits to the results of hydrodynamical simulations. We assume $\beta = 2.85$, as suggested by \cite{leinhardt2012}. $N_{\text{lr}} = 1$ and $N_{\text{slr}} = 2$ are the integer indexes representing LR e SLR, respectively. $C$ is a proportionality constant defined in \cite{leinhardt2012}. Once we have calculated $M_{\text{slr}}$, we can also compute the total mass in fragments as $M_{\text{frag}} = M_{\text{tot}}-M_{\text{lr}}-M_{\text{slr}}$.

\begin{table*}
\begin{tabular}{crrrcrrr}
\hline
$M_{\text{t}}$   & $M_{\text{p}}$   & $V_{\text{imp}}$   & \multirow{2}{*}{b}      & \multirow{2}{*}{Collisional Regime} &  \multicolumn{3}{c}{$M_{\text{frag}}$ ($M_{\oplus}$)}\\
($M_{\oplus}$) & ($M_{\oplus}$) & ($V_{\alpha,\text{esc}}$) &        &    &  \textsc{M-Flintstone}      & \textsc{LIPAD}   & \textsc{SyMBA}                           \\ \hline
3.7328 & 1.2850  & 1.1800     & 0.5787  & Partial Accretion        & 0.1218  & 0.0962  & 0.1218                   \\
2.3779 & 1.7084 & 1.8664   & 0.819      & Pure HnR      & 0.0     & 0.2404  & 0.0                  \\
3.1019 & 0.5349 & 2.2078   & 0.9165  & Erosive HnR      & 0.2169  & 0.1260  & 0.1702                  \\
5.7984 & 3.6118 & 4.2028   & 0.9428  & Super-catastrophic HnR  & 2.8996  & 2.9518  & 2.2840                     \\
4.1616 & 2.1033 & 11.0729  & 0.4275  & Super-catastrophic   & 6.2075  & 6.2647  & 6.2630                        \\
8.4263 & 4.0463 & 1.3960    & 0.7395 & Graze-and-merge  & 0.0     & 0.4519  & 0.0                             \\
1.6815 & 1.0191 & 5.1664   & 0.509  & Partial Erosion & 2.0510  & 1.2879  & 2.6999                           \\ \hline
 &  &    &   &  &  \multicolumn{3}{c}{Total ($M_{\oplus}$)}\\
 &  &    &   &  & 11.4968  & 11.4189  & 11.5389                           \\ \hline
\end{tabular}
\caption{\label{tab:colls} Comparing the outcome of selected collisions in \textsc{Mercury-Flintstone}, \textsc{LIPAD}, and \textsc{SyMBA}. Each line represents one particular collision. From left to right the columns show target mass, projectile mass, impact velocity normalised by the escape velocity, impact parameter, total mass in fragments produced by each code, and collisional regime following~\protect\cite{leinhardt2012}. The collisional algorithm included in \textsc{LIPAD} is based on the results from~\protect\cite{benzasphaug99} and that from \textsc{SyMBA} is described in~\protect\cite{chambers13}. \textsc{SyMBA} and \textsc{LIPAD} follow slightly different terminologies to refer to some collisional regimes relative to those used in this paper. Note also that for each collision the quantities $V_{\text{imp}}$  and $b$ are calculated in \textsc{Mercury-Flintstone} when the bodies collide. These quantities may be slightly different from those calculated in \textsc{LIPAD} and \textsc{SyMBA}, because these codes uses different numerical integrators and slightly different algorithms to find the very instant  when the bodies ``touch'' in a collision. This mostly accounts for the observed differences between the codes. The goal of this table is only demonstrate that the codes agree qualitatively well in terms of the total fragmented mass produced in different collisional regimes.}
\end{table*}

We calculated the mass distribution of fragments following the prescription (fits) provided in  \cite{leinhardt2012}. Due to computational limitations, we truncated the minimum fragment mass to $M_{\text{min}}$, to avoid an extremely large number fragments in our N-body simulations. After a series of test-simulations, we found that  $M_{\text{min}} = 10~M_{\text{Ceres}}$ provides a good compromise between the typical number of fragments created and the cpu-time required to complete our simulations.  Each simulation performed here requires between 2000 and 5000 CPU-hours on a single processor of a Xeon E5-2630 v4 processor at 2.2~GHz. 

The number of fragments with mass between $M_{\text{slr}}$ and $M_{\text{min}}$ is calculated as
\begin{equation}
N(M_{\text{slr}},M_{\text{min}}) = \frac{C}{\beta 2^{\beta }}\left ( M^{-\frac{\beta}{3} }_{\text{min}} -M^{-\frac{\beta}{3} }_{\text{slr}}\right )\left (\frac{4\pi \rho   }{3}\right )^{\frac{\beta}{3}},
\end{equation}
where $\rho$ represents the fragments' bulk density. The mass of the n-th fragment is iteratively calculated via the following equation
\begin{equation}
M_{\text{n}} = \frac{4\pi \rho }{3}\left (\frac{3(3-\beta )}{4\pi \rho C}M_{\text{rem}}  \right )^{3/(3-\beta )},
\label{eq:tailfrag}
\end{equation}
where $M_{\text{rem}}$ is the remaining mass in fragments after each iteration. When $M_{\text{n}} \leq M_{\text{min}}$, the remaining mass in fragments is distributed to (one or more) equal mass fragments with mass set equal to $M_{\text{min}}$. If at the very last  iteration, $M_{\text{rem}}\leq M_{\text{min}}$, we add the remaining mass to the previous fragment (iteration n-1) to ensure mass conservation rather than creating a new fragment. Our algorithm yields linear momentum conservation in collisions within a margin of $\sim$5\%. For collisions where $M_{\text{frag}} < M_{\text{min}}$, the collision is transferred to the perfect merge or graze-and-merge regimes.

We resolve collisions by placing the LR at collision's center of mass. The SLR and fragments are placed at equidistant positions from each other on a circumference located at the plane of the impact and centered at LR. The circumference radius corresponds to 1 hill radius of the total mass ($M_{\text{tot}}$). We assume that this circumference can accommodate up to 500 fragments, in order to prevent fragments from having ``overlapping'' starting positions. If more than 500 fragments are produced in a single collision we create another circumference with radius 10\% larger than the original one. We repeat this procedure until we can distributed all fragments around the body LR. We calculate the velocities of the LR, SLR and fragments following \cite{leinhardt2012}. The collisional model implemented here  does not take into account the possible pre-impact spin rates of colliding bodies. In a first approximation, the pre-impact spin may act reducing the disruption criteria (critical impact energy to lead to fragmentation), which may increase the fragmented mass. In our simulations, we assume that colliding bodies are not spinning.

We have compared our algorithm to solve collisions with those  used in   \textsc{LIPAD}~\citep{levisonetal12} and a modified version of \textsc{SyMBA}~\citep{duncanetal98,poon19,scoraetal2020}. We present these results in Table \ref{tab:colls}. Each line in Table \ref{tab:colls} represents one  collision event. From left-to-right  the columns show the masses of the target and projectile, impact velocity, impact parameter, total fragmented mass, and collisional regime. Table \ref{tab:colls}  shows that, for all collisions, the fragmented mass produced in \textsc{Mercury-Flintstone} agrees within a margin of $<$25\% (in most cases the difference is only a few percent) with that of one of the two codes. In our model, the fragment population is distributed following the mass/size distribution and velocities predicted by the results of hydrodynamical simulations~\citep{leinhardt2012}.  \cite{poon19} and \cite{scoraetal2020} follow \cite{chambers13}, and assume that the fragment population consist of equal mass/size objects. In \textsc{LIPAD}, fragments are distributed following the algorithms described in \cite{benzasphaug99} and \cite{morbidellietal09}. Although these codes invoke slightly different prescriptions for the distribution of fragments, we will discuss later that our results support qualitatively the main conclusions from previous studies using  \textsc{SyMBA} and \textsc{LIPAD}, and applied in different contexts.

\section{Simulations}\label{sec:simulations}

We have performed 100 N-body numerical simulations modelling the formation of super-Earths considering the effects of pebble accretion and planet-disk gravitational interactions. Our simulations come in two groups. The first group consists of 50 simulations where we neglect the effects of collisional fragmentation, i.e, collisions always result in perfect accretion and linear momentum conservation. We refer to this group of simulations as ``perfect set''. The second group consists of 50 simulations where collisions may also result in imperfect accretion, as described before. We refer to this group as ``imperfect set''. The prescriptions of type-I migration, gas tidal damping, pebble drift, pebble accretion, and gas disk models invoked in this paper are detailed described in~\cite{lambrechtsetal19} and~\cite{izidoro2019formation}. 

Our numerical simulations start from initial conditions equivalent to those of Model-I of~\cite{izidoro2019formation}. We have chosen this model in light of its success in matching the period ratio distribution of observations. Protoplanetary embryos are initially distributed between 0.7 and 20~au with masses randomly selected between 0.005 and 0.015~M$_{\oplus}$. Our simulations start when the gas disk is 3~Myr old (t$_{\rm start}=3$~Myr). Protoplanetary embryos are allowed to grow via accretion of drifting pebbles and collisions. The integrated pebble flux in our simulations -- from 3 to 5 Myr -- yields $\sim 150~$M$_{\oplus}$ in pebbles. The final masses of planets growing in simulations modelling pebble accretion strongly depends on the available pebble flux~\citep{lambrechtsetal19,izidoro2019formation,bitschetal19}. If the integrated pebble flux is very low (e.g. $\sim 40$~M$_{\oplus}$), planetary embryos tend to grow at most to Moon/Mars-mass planets~\citep{lambrechtsetal19,izidoroetal21} whereas a very high pebble flux (e.g. $\sim 300$~M$_{\oplus}$) tend to produce gas giant planets~\citep{bitschetal19}. The pebble flux chosen to conduct our simulations is intentionally designed to produce super-Earths mass planets as shown by ~\cite{izidoro2019formation}. The gas disk dissipates at 5~Myr and we extend our simulations up to 50~Myr in a gas-free scenario. We do not model gas accretion onto planetary embryos in our simulations. Our super-Earths are assumed to have a bulk density of 2~${\rm g/cm^3}$. This same density is used to compute the critical Q$'^*_{\text{RD}}$ in our collisional algorithm.

We do not directly compare the results of our simulations with observations in this paper. This analysis is conducted in \citep{izidoro2019formation}. We show, however, that the results of this paper support the main findings and conclusions of \cite{izidoro2019formation}. In the next section we show the dynamical evolution of growing planets in two representative simulations of the perfect and imperfect sets.

\subsection{Dynamical Evolution} \label{sec:example}

\noindent{
\begin{figure*}
	\includegraphics[width=0.7\paperwidth]{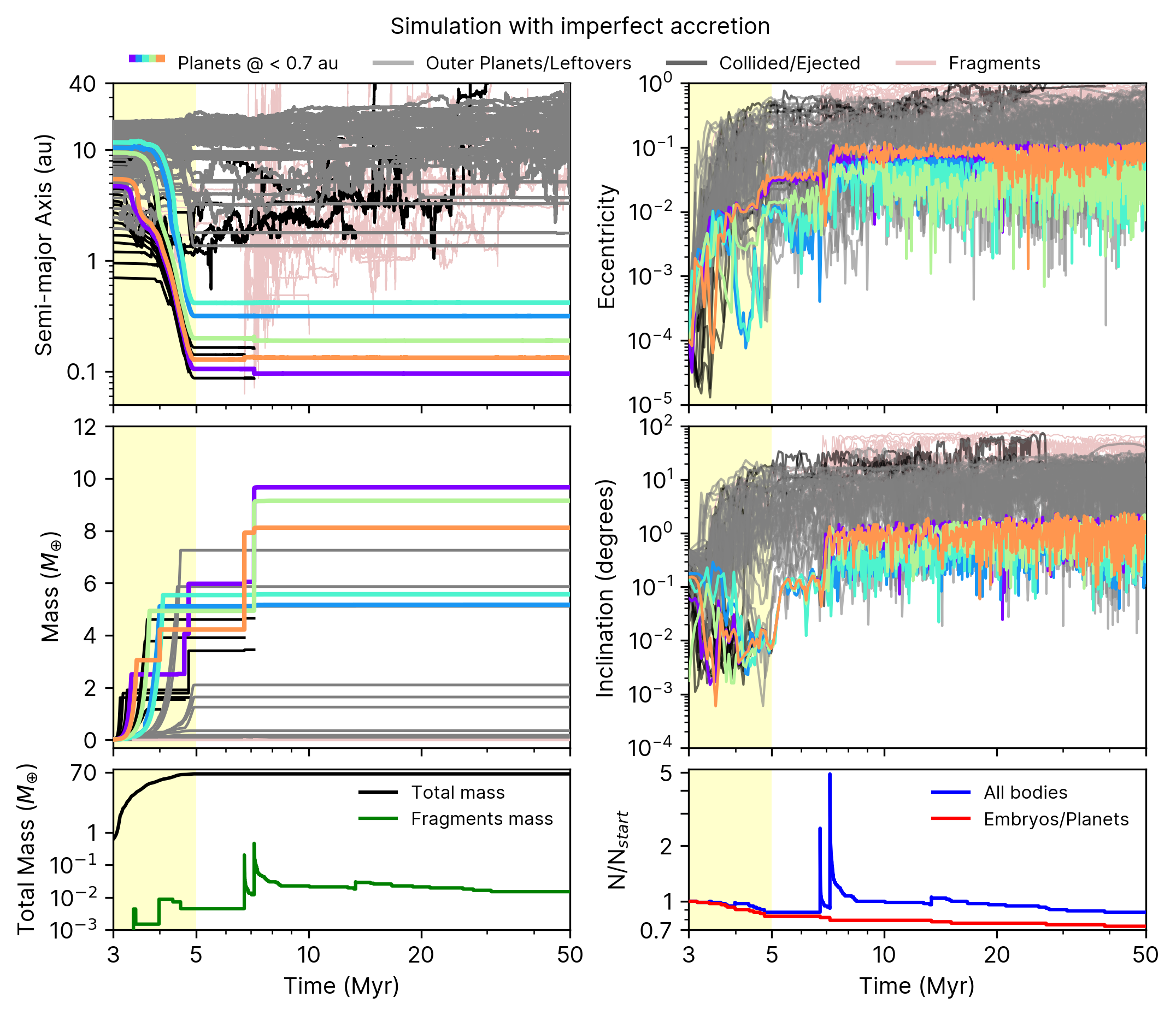}
    \caption{Growth and dynamical evolution of planetary embryos in a simulation where collisions may result in imperfect accretion. The light-yellow background marks the gas disk phase. Colourful lines illustrates the planets with final orbits inside 0.7 AU. The grey lines shows the planets or embryos leftovers with final orbits outside 0.7 AU. Black lines denotes collided/ejected planets/embryos. The light red lines show fragments produced in collisions.  {\bf Top left:} Temporal evolution of semi-major axis. {\bf Top right:} Evolution of orbital eccentricity. {\bf Middle left:} Temporal evolution of planetary mass. {\bf Middle right:} Evolution of orbital inclination. {\bf Bottom left:} Evolution of the total system mass (black line) and total fragment mass (green line). {\bf Bottom right:} The blue line shows the total number of bodies in simulation and red line shows the number of embryos/planets only. These number are normalized by the initial number of bodies in the simulation.}
    \label{fig:frag_compare}
\end{figure*}
}

Figure \ref{fig:frag_compare} shows the dynamical evolution of growing planets in a simulation of the imperfect set. This simulation starts with 71 protoplanetary embryos carrying an initial total mass of 0.6~$M_{\oplus}$ (bottom panel of Figure \ref{fig:frag_compare}). During the gas disk phase planets grow by pebble accretion and collisions. As they grow to sufficiently large masses ($ \gtrsim 0.1~M_{\oplus}$) they start to migrate (top-left panel of Figure \ref{fig:frag_compare}). More massive protoplanetary embryos scatter less massive ones on eccentric and inclined orbits (grey and black curves in the top-right and middle-right panels of Figure \ref{fig:frag_compare}). Collisions during the gas disk phase also lead to fragmentation (bottom-left panel of Figure \ref{fig:frag_compare}), although to some limited degree. At the end of the gas disk phase a chain of planets anchored at the disk inner edge set at 0.1~au forms (top-left panel of Figure \ref{fig:frag_compare}), with the most massive protoplanets having masses between 1 and 7~$M_{\oplus}$ (middle-left panel of Figure \ref{fig:frag_compare}). At the end of the gas disk phase the total mass carried by protoplanets is about $\sim$63~$M_{\oplus}$. After the gas disk dispersal the resonant chain becomes dynamically unstable, at 7~Myr. The dynamical instability promote collisions and fragmentation. Fragments that survive for more than 1000 years after their creation are shown via light red lines in Figure \ref{fig:frag_compare}. The planet represented by the orange line is the outcome of a partial accretion collision occurred during the instability phase. In this particular event 275 fragments were produced. The planet represented by the purple-line is also result of a grazing and merge collision during the instability phase. The green line shows a planet produced in a partial accretion accretion where 582 fragments were produced. The total mass in fragments produced during the instability phase is about 0.64~$M_{\oplus}$. Fragments produced during impacts tend to be rapidly accreted and scattered by protoplanets as shown by the bottom panels of Figure \ref{fig:frag_compare}. At the end of the simulation 5 planets survive inside 0.7~au with masses between 4 and 10~$M_{\oplus}$. 

\noindent{
\begin{figure*}
	\includegraphics[width=0.7\paperwidth]{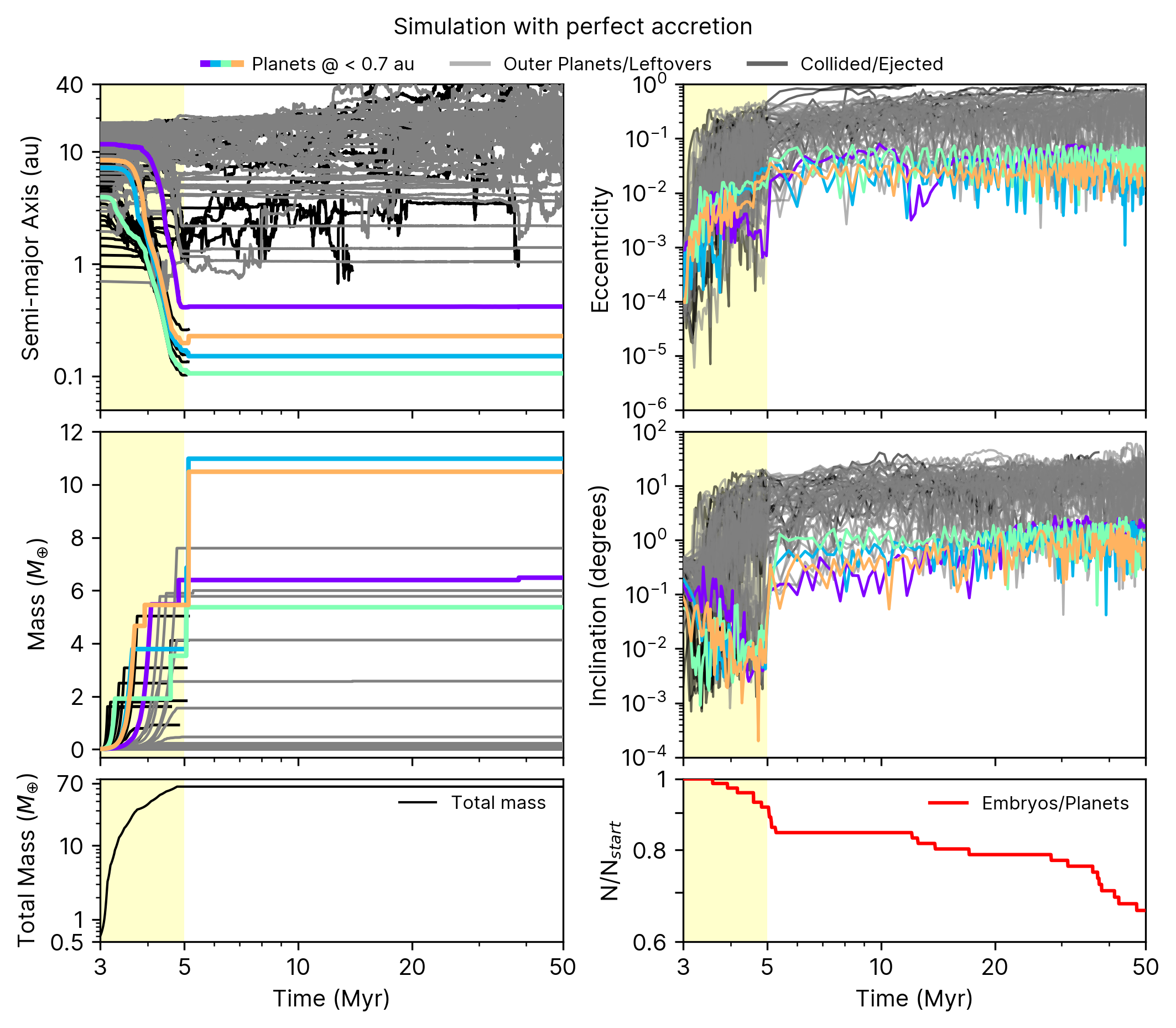}
    \caption{Same as Figure \ref{fig:frag_compare}, but for a simulation modelling collisions using the perfect accretion approach.}
    \label{fig:no_frag_compare}
\end{figure*}
}

Figure \ref{fig:no_frag_compare} shows the dynamical evolution of protoplanets in a simulation of perfect accretion set. This simulation starts from the same set of initial conditions (same mass and orbital elements for all protoplanets) and setup of that of Figure \ref{fig:frag_compare}. Although the overall dynamical evolution of the system in Figure \ref{fig:frag_compare} is similar to that of Figure \ref{fig:no_frag_compare}, the chaotic nature of planet formation simulations prevent us from directly comparing individual systems. In order to effectively measure the effects of imperfect accretion on the BTC scenario we need to invoke a statistical analysis. Note also that Figures \ref{fig:frag_compare} and \ref{fig:no_frag_compare} represent systems where the resonant chains become dynamically unstable after the gas disk dispersal (the ``unstable'' systems of \cite{izidoro2019formation}). We will discuss the fraction of unstable and stable systems in our simulations in Section 4.

\section{Statistical analysis}\label{sec:stat_analysis}

In this section we compare the dynamical architecture of planetary systems produced in our imperfect and perfect accretion sets of simulations using an statistical approach. In order to infer the role of fragmentation during and after the gas disk dispersal we divide our analysis in two steps. We first compare the imperfect and perfect set of simulations using the dynamical architecture of planetary systems produced at the end of the gas disk phase (5~Myr). Finally, we repeat this analysis comparing their dynamical state at the end of our simulations. We use the period ratio of adjacent planet pairs, planets mutual spacing, orbital eccentricity, orbital inclination, mass and planet multiplicity distributions as metrics to discuss our results.

\subsection{Dynamical architecture at the end of the gas disk phase}\label{subsec:gasdiskphase}

Following \cite{izidoro2019formation} we apply cutoffs in our data when conducting our statistical analysis. As we are particularly interested in the formation of hot super-Earths, only planets more massive than 1$M_{\oplus}$ and inside 0.7~au are accounted for in our analysis. These cutoffs are motivated by the high completeness of the Kepler-catalogue for planets larger than the Earth and with orbital period shorter than 200~days~\citep{petigura2013plateau} (for a solar-mass star a planet at 0.7~au has orbital period of about 200 days). All the same, we have verified that our results do not change qualitatively when we relax these thresholds and include all planets inside 2~au and with masses larger than 0.5$M_{\oplus}$.

Figure \ref{fig:gasdiskall} shows the period ratio of adjacent planet pairs, planets spacing, orbital eccentricity, orbital inclination, mass, and number of planets distributions in our simulations at the end of the gas disk phase (5~Myr). In agreement with previous studies \citep{izidoro2017breaking,ogiharaetal18,carreraetal19}, at the end of the gas disk phase planets are found in chain of mean motion resonances (top-left panel of Figure \ref{fig:gasdiskall}) and in low eccentricity and inclination orbits (middle-panels of Figure \ref{fig:gasdiskall}). More than 50\% of the planets formed in these simulations have masses between 2 and 6 M$_{\oplus}$. Both imperfect and perfect sets of simulations show remarkably similar distributions.  We have used KS-tests to compare the respective distributions of the imperfect and perfect sets of simulations shown in Figure \ref{fig:gasdiskall}. The respective p-values are shown inside each panel. By assuming that a p-value larger than 10\% implies that we can not reject the null hypothesis that both set are drawn from the same distribution we conclude that the imperfect and perfect set of simulations lead to statistically indistinguishable results. 

In the next section we show the impact velocity and geometry of collisions in our simulations and quantify the fraction of collisions falling in each collisional regime. 

\noindent{
\begin{figure*}
	\includegraphics[width=.83\paperwidth]{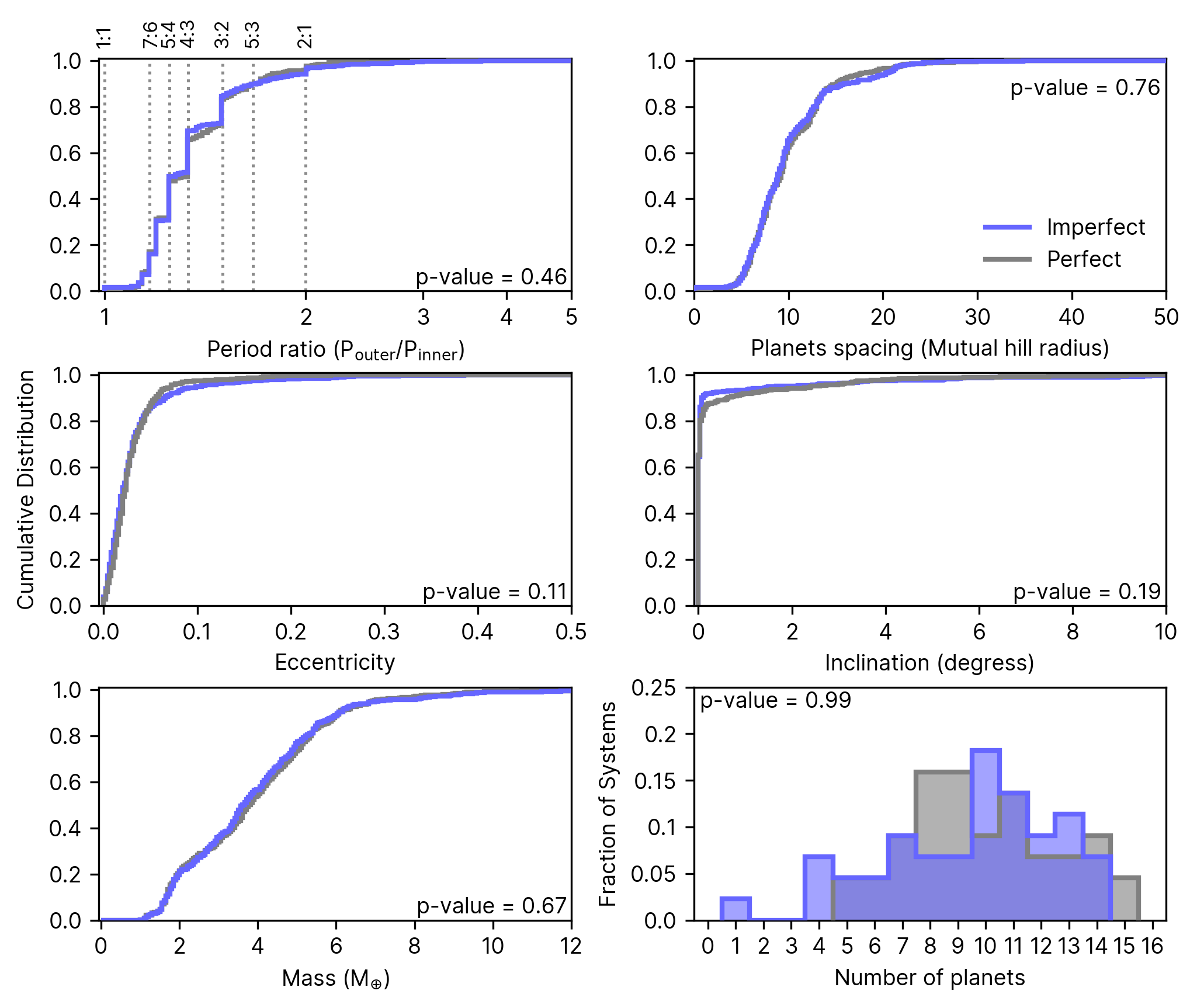}
    \caption{Dynamical architecture of planetary systems at the end of gas disk phase for planets/embryos that ended within 0.7~AU with masses greater than 1~$M_{\oplus}$. Blue lines denotes imperfect accretion set simulations and grey lines represent the simulations from perfect accretion set. Each panel also shows the p-value from K-S test comparing the respective distributions of the imperfect and perfect sets. {\bf Top left:} Cumulative distribution for the period ratio between adjacent planet pairs. Vertical dotted lines illustrate ratios values for possibles first-order mean motion resonances and co-orbital planet pairs. {\bf Top right:} Spacing between planet pairs given in terms of the mutual hill radius. {\bf Middle left/right:} Cumulative distribution for planets orbital eccentricity/inclination. {\bf Bottom left:} Cumulative distribution for planetary mass. {\bf Bottom right:} Multiplicity of planets in systems versus the fraction of simulations from imperfect and perfect sets. In some systems, the onset of dynamical instability phase takes place just before the end of the gas disk phase, typically during the last few $\sim$100~Kyr. We have verified that these same distributions at 4.75~Myr are only slightly different from those at 5~Myr. More importantly, the KS-tests for all distributions still support our conclusions that the distributions of the perfect and imperfect sets are statistically the same.}
    \label{fig:gasdiskall}
\end{figure*}
}

\subsection{Collisions during the gas disk phase}

Figure \ref{fig:gasdiskcoll} shows all collisions that occurred during the gas disk phase in the imperfect set of simulations. Colour-coding is used to show their respective regime as described in Section 2. The left panel shows the impact geometry of all collisions. The top-right panel of Figure \ref{fig:gasdiskcoll} shows the frequency and fragmentation efficiency of each regime. Perfect merge and graze-and-merge regimes corresponds to about 44.5\% of all collisions taking place during the gas disk phase. About 40\% of the collisions fall into the partial accretion regime. About 16\% of the cases are erosive, pure, and super-catastrophic hit-and-run impacts.  Although the fraction of collisions in the partial-accretion regime is considerably high the top-right panel of Figure \ref{fig:gasdiskcoll}  shows that, on average, the total mass in fragments produced in these collisions corresponds to a few percent of the total colliding mass. This is also true for collisions in the super-catastrophic and erosive hit-and-run. One of the most energetic collisions observed in our simulations occurred between two Mercury-mass fragments (by-products of an anterior collision) that collided at velocities of  $\sim$17~$V_{\alpha,\text{esc}}$. In this particular case a super-catastrophic collision occurred with more more than 90\% of the colliding mass being fragmented. Note, however, that super-catastrophic and partial erosion collisions are extremely rare.

\noindent{\begin{figure*}
	\includegraphics[width=.73\paperwidth]{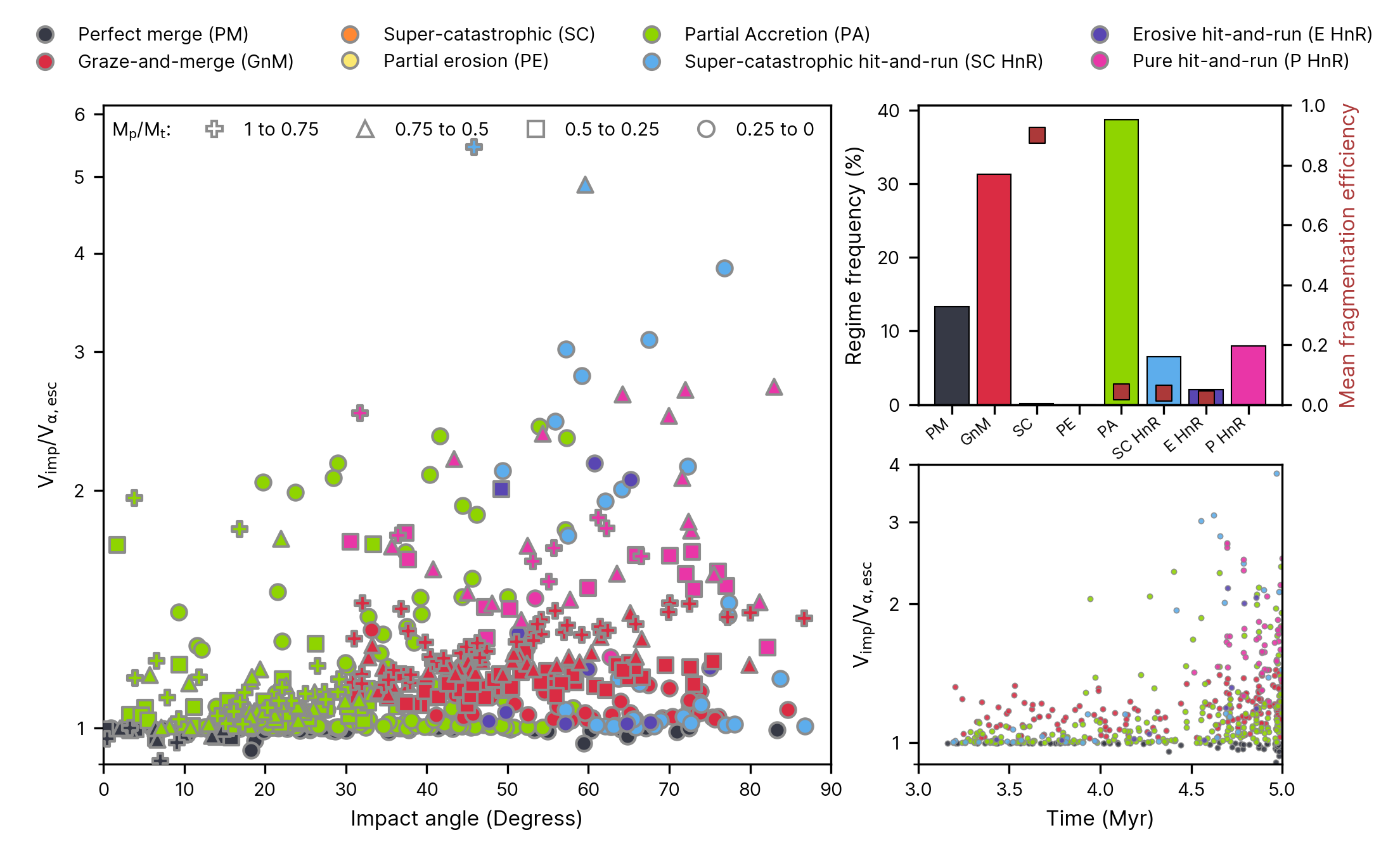}
    \caption{{\bf Left panel} Impact geometry and velocity in collisions taking place during the gas disk phase in the imperfect set of simulations. Colour-coded symbols represent different collisional regimes. Each symbol shape denotes a mass-ratio interval for the collision. The impact velocity is normalised by the surface escape velocity of the interacting total mass. {\bf Top right panel} shows the frequency (bars plot) and mean fragmentation efficiency (squares plot) of each regime. {\bf Bottom right panel} shows the impact velocities in function of time.}
    \label{fig:gasdiskcoll}
\end{figure*}}

We have found that only 5\% of all collisions takes place at impact velocities higher than 2~$V_{\alpha,\text{esc}}$. This is not a surprising result given that the gas helps to damp the orbital eccentricities and inclination of planetary embryos during the gas disk phase. In fact, the bottom-right panel of Figure \ref{fig:gasdiskcoll} shows that collisions taking place early in the disk phase tend to show lower impact velocities than those happening later. The reason for this is two fold. The impact velocity scales with the surface escape speed of the colliding bodies and their respective random velocities (their velocities before they become bounded). During the early stages of the disk, planetary embryos are less massive so their escape velocities are lower. During the early stages the gas disk is also more dense, and damping of orbital inclination and eccentricity is more efficient. As the disk dissipates and bodies become more massive their impact velocities tend to increase, yet not up to the point where the effects of imperfect accretion becomes sufficiently important.

We have found that  most fragments created during the gas disk phase have  masses smaller than Moon-mass. Fragments more massive than the Moon are only produced when the gas disk is almost fully dissipated and, consequently, they have virtually no time to accrete pebbles and grow via pebble accretion. In our model, we neglect pebble accretion onto fragments with masses lower than Moon-mass \cite[e.g.][]{johansenlambrechts17}. We have verified that $\sim$70\% of the fragments created during the gas disk phase are quickly reacreted by larger protoplanetary embryos in timescales of  $\leq$50~kyr, so even if accounted for, they would have not time to grow significantly via pebble accretion. Fragments that survive longer timescales have typically eccentric/inclined orbits so pebble accretion on these fragments would be also  negligible~\citep[e.g.][]{levisonetal15,johansenlambrechts17}. As fragments produced during the early disk phase are generally too small to accrete pebbles (or dynamically too hot) and fragments produced during the late disk phase  do not have enough time to accrete significant mass in pebbles before the gas dissipates, fragments do not significantly increase the final mass of the planetary system compared to the perfect case (e.g. bottom-left panels of Figures \ref{fig:frag_compare} and \ref{fig:no_frag_compare}). If pebble accretion on fragments was efficient, one could expect to have either a larger number of (potentially Earth-mass) planets in the system or more massive planets in the imperfect accretion case compared to the perfect case, which is clearly also not the case in Figure \ref{fig:gasdiskall}. We can conclude that during the gas disk phase imperfect accretion have a negligible effect on the final dynamical architecture of super-Earth systems forming via pebble accretion and migration due to the limited effect of fragmentation.

\subsection{Final dynamical architecture}\label{subsec:longtermevolution}

In this section we compare the dynamical architecture of planetary system in the perfect and imperfect set of simulations at the end of our simulations, namely at 50~Myr. We have found that
$\sim$10\% of the resonant chains produced in the imperfect set remained dynamically stable until the end of our simulations, at 50~Myr. 4\% of the resonant chains of the perfect set remained stable. Due to the small number of ``stable'' systems, we cannot firmly conclude that the difference in the fraction of ``stable'' systems between the two sets is statically significant. A larger set of simulations would be desirable to provide a statically more robust conclusion.
However, the low fraction of stable systems in these simulations is qualitatively consistent with the results of \cite{izidoro2019formation}.

Following \cite{izidoro2017breaking,izidoro2019formation}, we divided our planetary systems in two categories, the {\it unstable} and {\it stable} systems. Figure \ref{fig:finalall} shows the orbital architecture of {\it unstable} and {\it stable} planetary systems of the perfect and imperfect sets of simulations after 50 Myr of integration. As for Figure \ref{fig:gasdiskall}, Figure \ref{fig:finalall} shows the period ratio, planet mutual spacing, orbital eccentricity, orbital inclination, mass, and number of planets distributions. Figure \ref{fig:finalall} shows that -- similar to what we have found at the end of the gas disk phase -- the overall dynamical architecture of planetary systems in the perfect and imperfect set of simulations are remarkably similar. As expected, planets in {\it stable} systems tend to have compact and resonant orbits whereas unstable system have more spread and dynamically excited orbits. Figure \ref{fig:finalall} also shows in each panel the p-values calculated from KS-tests when we compare the respective distributions of our imperfect and perfect set of simulations. We do not apply KS-tests to compare stable systems due to the limited number of stable system in the sample. As one can see in all panels of Figure \ref{fig:finalall}, the distributions of stable systems in the perfect and imperfect sets do show some differences, but these are mostly due to small number statistics. The p-values calculated from our KS-tests confirm that the distributions of unstable systems at the end of simulations are statistically the same in the perfect and imperfect sets.

\noindent{
\begin{figure*}
	\includegraphics[width=.73\paperwidth]{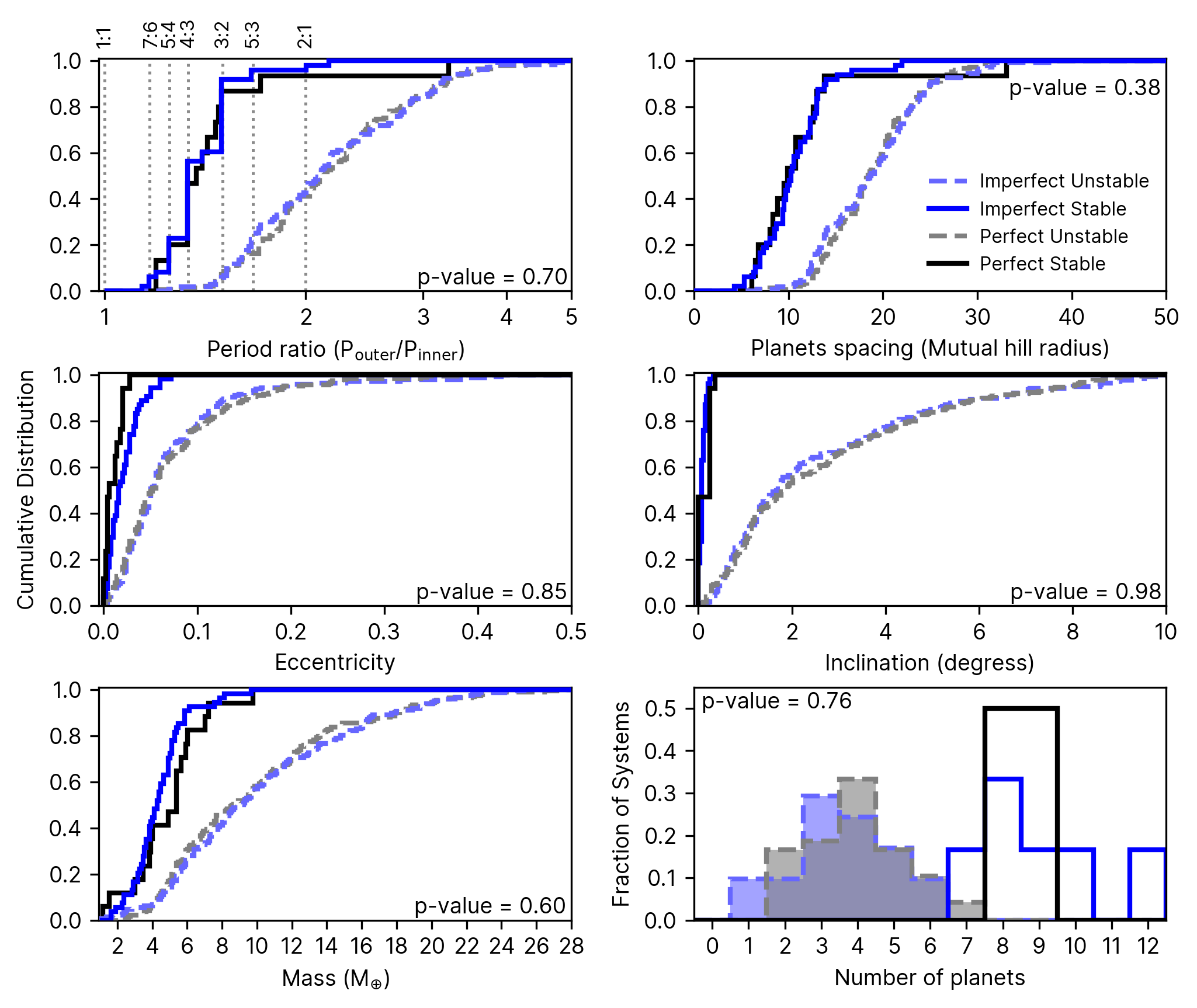}
    \caption{Final architecture of {\it unstable} (dashed lines) and {\it stable} (solid lines) planetary systems at 50~Myr. Panels shows the p-value from K-S test comparing the distributions from {\it unstable} systems of imperfect and perfect sets. Only planets/embryos with semi-major axis smaller than 0.7~AU and masses greater than 1~M$_{\oplus}$ are taken into account. Blue colours and grey-scale colours illustrates the simulations from imperfect and perfect accretion sets, respectively. Panels are the same as those in Figure \ref{fig:gasdiskall}.}
    \label{fig:finalall}
\end{figure*}
}

\noindent{
\begin{figure*}
	\includegraphics[width=.73\paperwidth]{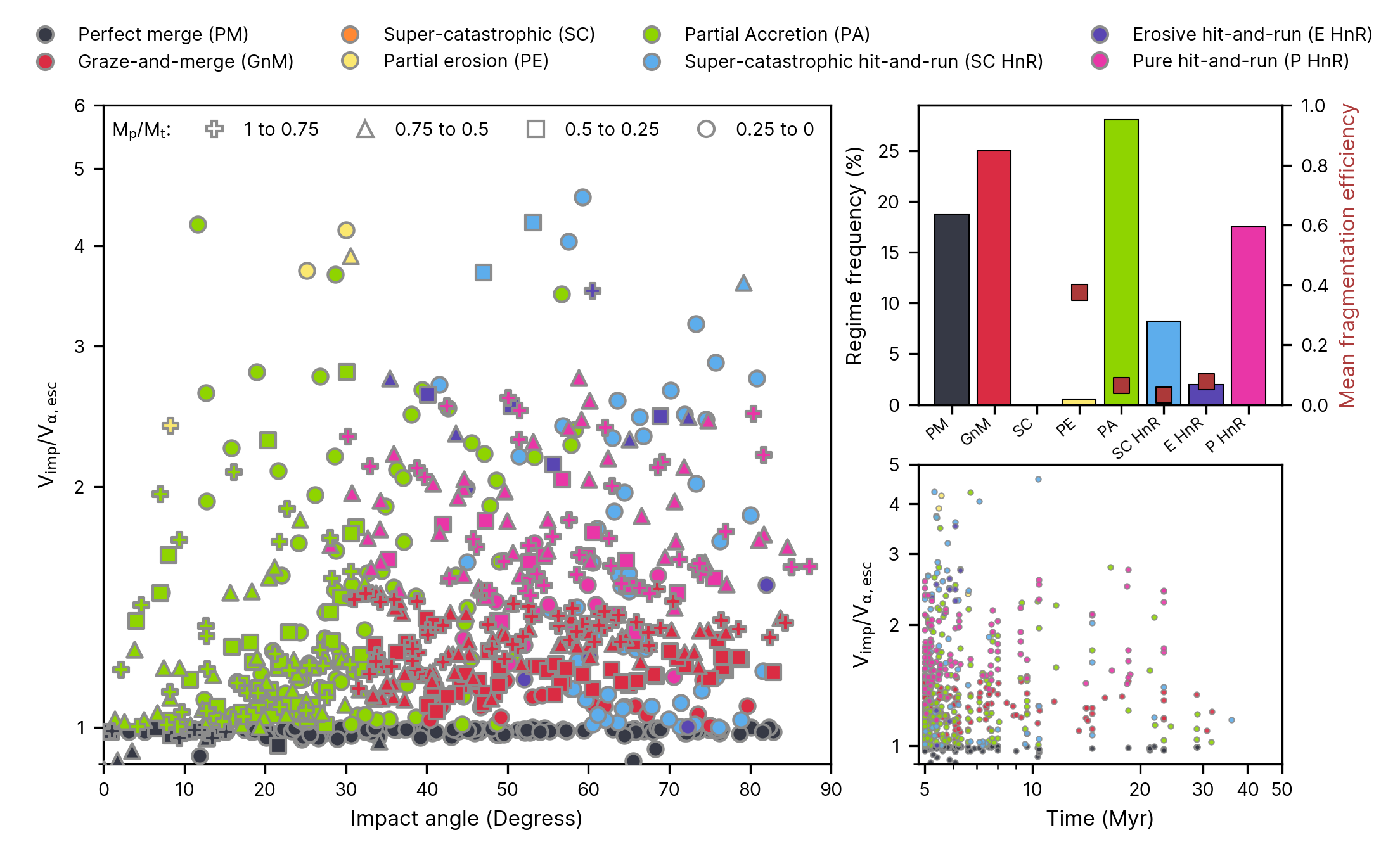}
    \caption{Same as Figure \ref{fig:gasdiskcoll}, but for collisions taking place after the gas disk dispersal, from 5~Myr to 50~Myr.}
    \label{fig:colltypes}
\end{figure*}
}

\subsection{Collisions after the gas disk phase}

\noindent{
\begin{figure}
	\includegraphics[scale=0.86]{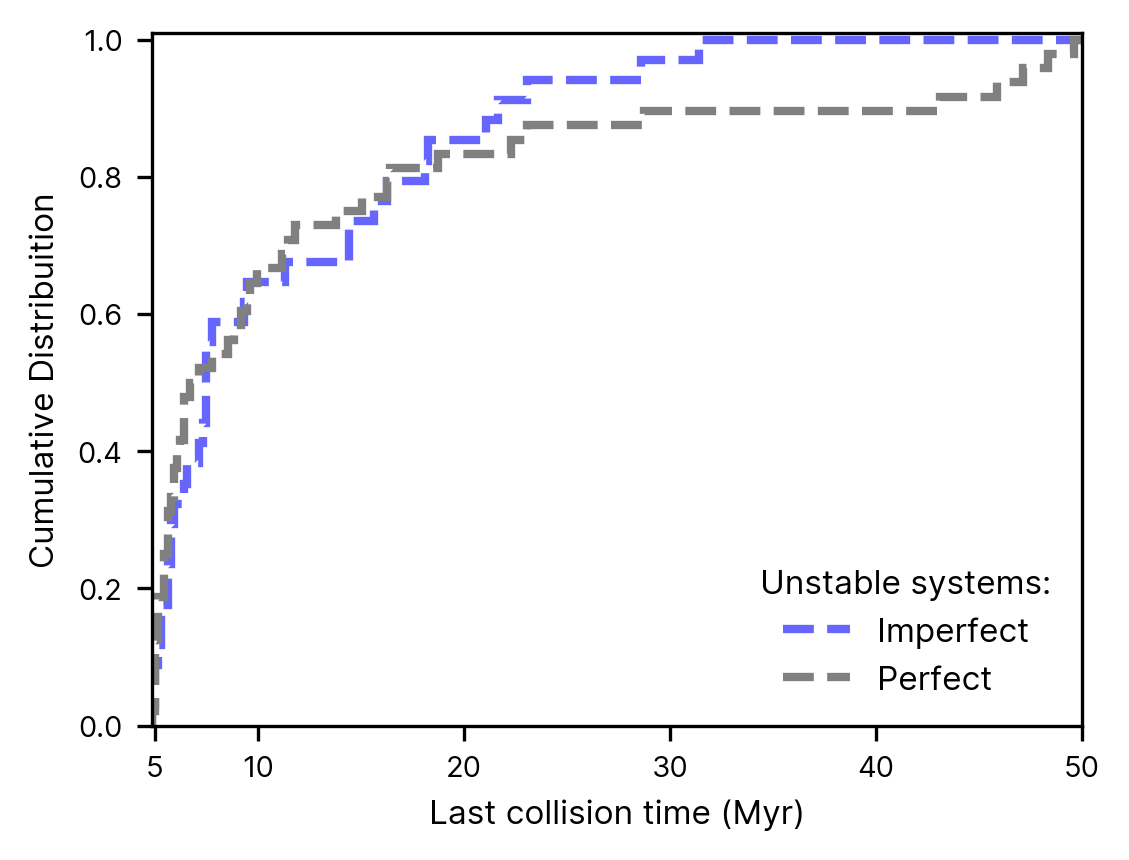}
    \caption{Cumulative distribution of the last collision  event taking place after the gas disk phase. We define a giant collision as those involving planetary objects with masses larger than 1~M$_{\oplus}$ and orbiting within 0.7~AU from the star (we have verified that this figure does not change qualitatively when we include all planets  with masses $\geq$0.5~M$_{\oplus}$ and within 2~au). Blue dashed line shows imperfect accretion set simulations, while gray dashed line shows the perfect accretion set simulations.}
    \label{fig:lastcolltime}
\end{figure}
}

Figure \ref{fig:colltypes} shows all collisions that occurred after the gas disk phase in the imperfect set of simulations. As in Figure \ref{fig:gasdiskcoll}, colour-coding is used to represent different regimes. The left panel shows the geometry of collisions. The top-right panel of Figure \ref{fig:colltypes} shows the frequency and mean fragmentation efficiency of each regime. If we compare the results of  Figures \ref{fig:gasdiskcoll} and Figure \ref{fig:colltypes} it is evident that collisions after the gas disk dispersal tend to happen at higher impact velocities, which is expected due to the lack gas tidal damping. About 44\% of all collisions fall in the perfect merge and graze-and-merge regimes, compared to 43\% of the collisions in the gas disk phase. 28\% and 17\% of the collisions corresponds to the partial accretion and pure hit-and-run regimes, respectively. 10\% of the collisions fall into the super-catastrophic and erosive hit-and-run regimes. Partial erosion corresponds to only 0.5\% of the all collisions taking place after gas dispersal. Except for the collisions in the partial erosion regimes, the fraction of the colliding mass fragmented in non-accretionary events tends to be lower than $\sim$10\%. Collisions in the partial erosion regime fragmented on average 40\% of the colliding mass. The right-bottom panel of Figure \ref{fig:colltypes} shows that most collisions taking place after the gas disk dispersal happens shortly after the gas is gone (first 5~Myr). This is consistent with the results of \cite{izidoro2017breaking} and \cite{izidoro2019formation}. Finally,  Figure \ref{fig:lastcolltime} shows that imperfect accretion has no dramatic impact on the timing of the last giant impact on super-Earths (see \cite{kokubogenda10} for a similar conclusion in the context of the solar system). 

Our results shows that the breaking the chains scenario typically leads to collisions that result in limited fragmentation. The total mass in fragments produced in typical collisions corresponds to less than 10\% of the total mass of the colliding planets. The fragmented mass tend to be rapidly accreted by the more massive remnants of the collision and proves to be inefficient in promoting residual planetesimal-driven migration or efficient damping of orbital inclination and eccentricities of close-in (see \cite{deiennoetal19} for a similar conclusion in the solar system context).

\section{Discussion}\label{sec:discussion}

We have showed that from a statistical point of view the effects of collisional fragmentation have negligible impact on the final dynamical architecture of hot super-Earths systems produced in the breaking the chains scenario. We can now discuss other possible effects of collisional fragmentation.

\noindent{
\begin{figure}
\centering
	\includegraphics[scale=0.8]{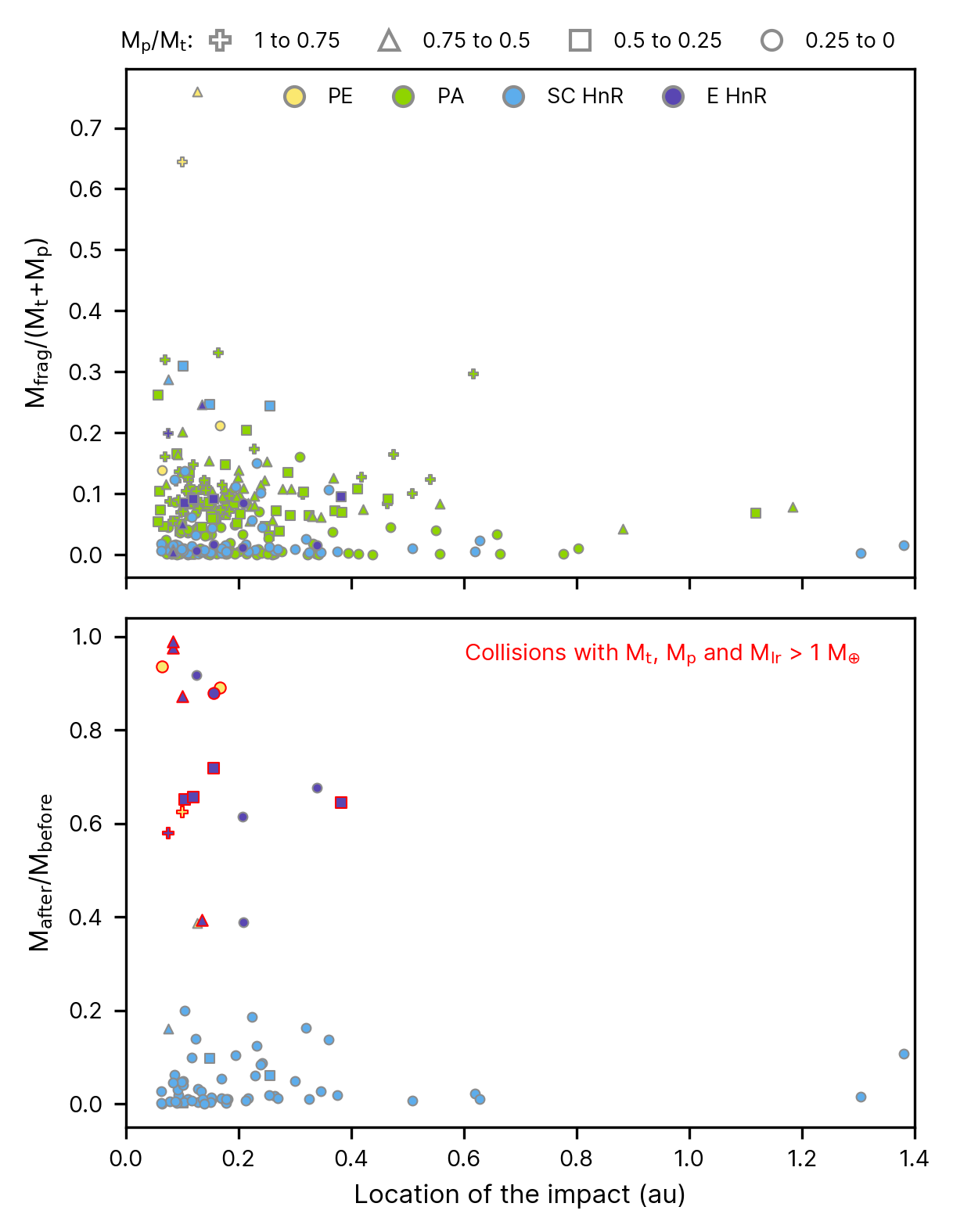}
    \caption{{\bf Top:} Total mass in fragments normalized by the mass of the colliding bodies (M$_{\textrm{frag}}$) in function of the location of the impact. Each data-point represent individual collisions taking place after gas disk dissipation.  {\bf Bottom:} Remaining mass in erosive collisions in function of the impact location. Each data point represents one specific eroded planet. In both panels, different symbols are used to represent a given mass-ratio interval (top) while the colours represent a specific collisional regime. Collisions highlighted in red are those where the surviving bodies have masses larger than 1~M$_{\oplus}$.}
    \label{fig:r_vs_fragmass}
\end{figure}
}

\subsection{Dust production during the instability phase}

The presence of dust disks around stars is typically inferred via  measurements of infrared excess due to thermal emission of circumstellar dust. Observations suggest that stars older than $\sim$10~Myr typically do not host hot dust disks~\citep{haisch01,wyatt08}. Although some stars older than $\sim$10~Myr do show dust disks, their disks tend to be significantly less massive than disks of much younger stars. This strongly suggests that most stars loose their disks in about 10~Myr. Dust around old disks could be remnants of conventional protoplanetary disks or the result of dust produced via collisions of planets/planetesimals after the dissipation of protoplanetary disks~\citep{wyatt08}. In this section we discuss the timing and amount of dust produced in our simulations and discuss our results in the context of disk observations.

Figure \ref{fig:finalall} shows that in our model resonant chains tend to become dynamically unstable soon after gas disk dissipation, which lead to orbital crossing and collisions. The instability phase is typically short because of short dynamical timescales close to the star. Figure \ref{fig:lastcolltime} shows that ~60\% of our systems have their last giant collision within 5~Myr after gas disk dissipation, and 90\% of the systems have the last giant collision within 20~Myr. In our simulations we do not directly model the fate of dust produced in collisions because we truncate the distribution of fragments to a minimal size of 10 Ceres mass. However we can still estimate  the amount of dust that would be produced in our collisions.

Following \cite{leinhardt2012}, an erosive collision between two hot super-Earths produce a population of fragments with differential size distribution given by $n(D)dD \propto D^{-(\beta +1)}dD$ , where $\beta = 2.85$. If we assume that fragments with sizes smaller than 1-meter will be converted to dust via collisional cascade we find that $\sim$25\% of the fragmented mass in our collisions will be converted into dust, where the total mass in fragments typically corresponds to $\sim$5\% of the colliding mass (M$_t+$M$_p$). If two typical super-Earths of 5~M$_{\oplus}$ collide in a typical event, the total mass produced in dust grains smaller than $<$1~m will be about 0.125~M$_{\oplus}$. Warm dust has been observed at distances from the star $\lesssim$1~au~\citep[e.g.][]{kennedywyatt12,thompsonetal19}. Estimates of the dust mass in disk older than 10~Myr yields dust reservoirs of $\sim$0.01 – 0.1~M$_{\oplus}$~\citep{wyatt08}. If for simplicity we neglect dust reaccretion onto the planets, and assume that typical resonant chains exhibits at least a few erosive impacts one can speculate that the breaking the chains scenario would produce an amount of dust that is at least marginally consistent with that observed around stars older than 10~Myr. Dust particles of 1~cm and density of 2~g/cm$^3$ at 0.2~au would have a typical lifetime of only 600~kyr, whereas particles at 0.7~au would have a typical lifetime of about 6~Myr~\citep{wyattwhipple1950}. The short duration of the instability phase in our simulations together with the expected short lifetime of dust grains in the innermost regions of the disk are also qualitatively aligned with disk observations. Warm dust observed inside 1~au around stars  older than than $\sim$100-10000~Myr~\citep[e.g.][]{thompsonetal19} may be: i) the result of late impacts between inner planets subsequent to the breaking the chain evolution; ii) the result of long-term stable systems  becoming eventually unstable; or iii) collisions of inner planets with leftover outer planetesimals/protoplanets scattered inward~\cite[e.g.][]{wyatt08}.

\subsection{Mantle erosive collisions and high bulk-density exoplanets}

Hot super-Earths are thought to be fully differentiated bodies composed of an iron core, rocky/icy mantle and a potential thin gaseous envelope. The bulk density of an observed planet may be estimated by combining different observations techniques as transit, radial velocity, transit timing variations, etc~\citep[e.g.][]{weissmarcy14,haddenlithwick17}. The bulk density of the large majority hot super-Earths remains poorly constrained but recent studies have suggested that up to $\sim$20\% of the observed innermost hot super-Earths may have bulk densities higher than that of Earth and consistent with that of Mercury~\citep{adibekyanetal21,schulzeetal21}. Mercury's mean density is 5.43~g/cm$^3$ and it has been proposed to be the outcome of an energetic mantle-stripping collision of two Mars-mass protoplanets~\citep{asphaugreufer14} during the formation of the terrestrial planets in the solar system. Mantle-stripping collisions may remove significant part of the planets' mantle potentially leaving behind a high density iron-rich core (CMF)~\citep{marcusetal09,asphaugreufer14}. In this section we investigate how common are mantle-stripping collisions  in our simulations. 

We first investigate how the occurrence of imperfect accretion events correlate with the heliocentric distance. The top panel of Figure \ref{fig:r_vs_fragmass} shows that the most erosive collisions tend to occur in the innermost regions of the system. Erosive impacts taking place inside 0.2~au fragment up to 60-70\% of the total colliding mass. Collisions taking place in the innermost regions of the disk tend to be more energetic because of higher orbital velocities in the inner regions of the disk and higher probability of collision with bodies at high eccentricity orbits. The bottom panel of Figure \ref{fig:r_vs_fragmass} shows the mass ratio of planetary bodies experiencing erosive collisions (collisions in the partial accretion regime are not shown) just after and before collision. We do not model the interior structure of planets in this work. Super-Earths are expected to have iron mass fractions varying between $\sim$10\% and 40\% and super-Mercuries varying between $\sim$50\% and $\sim$80\%~ \citep{adibekyanetal21}. We are particularly interested in collisions where from $\sim$20 to $\sim$80\% of the putative body mass is eroded during the impact and the remaining body have mass larger than 1~M$_{\oplus}$.  These collisions are the most likely to produce hot super-Earths with high CMF, i.e., super-Mercuries, if a sufficiently large fraction of the mantle is eroded.

We found 7 collision satisfying to these criteria, 6 falling in the erosive hit-N-run regime and one case in the partial erosion regime. We have found that these highly erosive events are noticeably rare and correspond to only 5\% of the impacts of bodies more massive than 1~M$_{\oplus}$. In reality, we have also found that all planetary objects produced in these impacts were subsequently  recreated by other more massive planets during the system evolution. Thus, it seems unlikely that the breaking the chains scenario can account for the apparent abundance of super-Earths with high CMF (high densities), the so called super-Mercuries~\cite[e.g.][]{adibekyanetal21}. The iron fraction of some observed inner planets may be higher than the standard picture because inward drifting iron rich pebbles evaporate and then recondense, increasing the iron fraction of the pebbles close to the iron evaporation front~\citep[e.g.][]{aguichineetal20}. This interpretation seems to be aligned with the results of interior models~\citep{Dornetal19} suggesting that in order to account for the composition of a specific set of super-Earths these planets have to form at special locations, where certain elements are less/more abundant in order to explain the planets' peculiar compositions.

\section{Conclusions}\label{sec:conclusion}

In this paper we have used N-body numerical simulations including the effects of planet-disk interaction and pebble accretion to revisit the breaking the chains scenario for the formation of super-Earths~\citep{izidoro2017breaking,izidoro2019formation,lambrechtsetal19,carreraetal19}. Previous simulation of the breaking the chains scenario have modelled collisions as perfect merging events. Here we compare the results of simulations following this traditional approach with the results of simulations where a more realistic treatment of collisions is considered. We have developed a modified version of the code \textsc{Mercury-Flintstone}~\citep{chambers1999hybrid,izidoro2019formation} implementing the algorithm of \cite{leinhardt2012} to resolve collisions which is calibrated from a suite of impact hydrodynamical simulations. We have performed two sets of 50 simulations starting from the very same initial conditions, with the only difference being the scheme adopted to resolve collisions. In one case, collisions are modelled as perfect merging events conserving mass and linear momentum (``perfect'' set). In the other case, collisions are modelled allowing imperfect accretion (``imperfect'' set). In our simulations, protoplanetary embryos grow from roughly Moon-mass to larger masses by accreting pebbles and collisional fragments in the disk. As planets grow they migrate inwards reaching the disk inner edge and forming long chains of mean motion resonances. The gas disk is assumed to dissipate at 5~Myr and our simulations are extended in a gas-free scenario up 50~Myr. After the gas disk dispersal a large fraction of our compact resonant chains become dynamically unstable. We refer to this as the breaking the chains scenario. We have compared the dynamical architecture of planetary systems produced in our two set of simulations at two different epochs, namely at the time of the gas disk dispersal and at the end of our simulations. We compared the period ratio, mutual spacing, orbital eccentricity, orbital inclinations, mass, and planet multiplicity distributions of the perfect and imperfect set of simulations and found that both approaches leads to  statistically equivalent results. This is true both when we compare our systems at the end of the gas disk phase and at 50~Myr, i.e,  after the instability phase. Although imperfect accretion events are very common, accounting for  more than 50\% of the collisions, we found that these collisions typically lead to limited fragmentation. Only $\sim$10\% of the system mass is fragmented during a typical “late instability phase”. In addition, most fragments are rapidly reaccreated rather than ejected from the system. Although applied in a different scenario, our results are qualitatively aligned to the results of previous studies modelling the formation of terrestrial planets~\citep{kokubogenda10,chambers13,walshlevison16,deiennoetal19,walshlevison19,clementetal19} and super-Earths including the effects of imperfect accretion~\citep{poon19,scoraetal2020}. Imperfect accretion has negligible effect on the formation and dynamical configuration of hot super-Earths systems. Future studies of the breaking the chains scenario should now focus on the role of giant impacts on sculpting the atmospheric envelopes of hot super-Earths. 

\section*{Acknowledgements}
We are very grateful to the referee, John Chambers, for comments and suggestions that helped to improve our paper. L.~E. is grateful to FAPESP for financial support through grants 19/02936-0, 20/07689-8 and 21/00628-6. A.~I. thanks CNPq (313998/2018-3) and FAPESP (16/19556-7; 16/12686-2) for support during the initial development of this work.  A.~I. also thanks NASA for support via grant 80NSSC18K0828 to Rajdeep Dasgupta, during preparation and submission of the work. B.~B., thanks the European Research Council (ERC Starting Grant 757448-PAMDORA) for their financial support. S.~N.~R. thanks the CNRS's PNP program for support. O.~C.~W. thanks Conselho Nacional de Desenvolvimento Científico e Tecnológico (CNPq) proc. 312813/2013-9 and FAPESP proc. 16/24561-0 for financial support. O.~C.~W and A.~I  thank the Brazilian Federal Agency for Support and Evaluation of Graduate Education (CAPES), in the scope of the Program CAPES-PrInt, process number 88887.310463/2018-00, International Cooperation Project number 3266.  \\

\section*{Data availability}
The data underlying this article will be shared on reasonable request to the corresponding author.




\bibliographystyle{mnras}
\bibliography{references} 





\bsp	
\label{lastpage}
\end{document}